\documentclass[twocolumn,preprintnumbers,tightenlines,nofootinbib,superscriptaddress]{revtex4-1}
\pdfoutput = 1

\usepackage{amssymb}
\usepackage{amsmath}
\usepackage{epsfig}
\usepackage{comment}
\usepackage{color}
\usepackage{hyperref}
\usepackage{cleveref}
\usepackage{multirow}
\usepackage{capt-of}
\usepackage{siunitx}
\usepackage{graphicx}
\usepackage{gensymb}
\usepackage{slashed}
\usepackage{tabularx}
\usepackage{enumitem}
\usepackage{multirow}
\usepackage{booktabs}
\usepackage{isotope}
\usepackage{xcolor}
\usepackage{bm}
\usepackage{array,mathtools}
\usepackage{tikz}

\usepackage[sort&compress]{natbib}

\newcommand{\su}[1]{\ensuremath{SU(#1)}}
\newcommand{\U}[1]{\ensuremath{U(#1)}}
\newcommand{\order}[1]{\ensuremath{\mathcal{O}(#1)}}

\newcommand{\gamp}{{\gamma '}}
\newcommand{\RH}{{\text{R}}}
\newcommand{\GeV}{\ensuremath{\,\text{GeV}}}
\newcommand{\MeV}{\ensuremath{\,\text{MeV}}}

\newcommand{\Eq}[1]{Eq.~\eqref{eq:#1}}
\newcommand{\prn}[1]{ \left(  #1 \right) }
\newcommand{\al}[1]{\begin{align} #1 \end{align}}

\begin{document}

\title{Asymmetric Matter\emph{s} from a Dark First-Order Phase Transition}

\author{Eleanor Hall}
\email{nellhall@berkeley.edu}
\affiliation{Department of Physics, University of California, Berkeley, CA 94720, USA}
\affiliation{Ernest Orlando Lawrence Berkeley National Laboratory, Berkeley, CA 94720, USA}

\author{Thomas Konstandin}
\email{thomas.konstandin@desy.de}
\affiliation{DESY, Notkestraße 85, D-22607 Hamburg, Germany}

\author{Robert McGehee}
\email{robertmcgehee@berkeley.edu}
\affiliation{Department of Physics, University of California, Berkeley, CA 94720, USA}
\affiliation{Ernest Orlando Lawrence Berkeley National Laboratory, Berkeley, CA 94720, USA}

\author{Hitoshi Murayama}
\email{hitoshi@berkeley.edu}
\affiliation{Department of Physics, University of California, Berkeley, CA 94720, USA}
\affiliation{Kavli Institute for the Physics and Mathematics of the
  Universe (WPI), University of Tokyo,
  Kashiwa 277-8583, Japan}
\affiliation{Ernest Orlando Lawrence Berkeley National Laboratory, Berkeley, CA 94720, USA}
\affiliation{DESY, Notkestraße 85, D-22607 Hamburg, Germany}

\begin{abstract} 
We introduce a model for matter{\it s}-genesis in which both the baryonic and dark matter asymmetries originate from a first-order phase transition in a dark sector with an $\su3\times\su2\times\U1$ gauge group and minimal matter content. In the simplest scenario, we predict that dark matter is a dark antineutron with mass either $m_{\bar{n}} = 1.36$ GeV or $m_{\bar{n}} = 1.63$ GeV. Alternatively, dark matter may be comprised of equal numbers of dark antiprotons and pions. This model, in either scenario, is highly discoverable through both dark matter direct detection and dark photon search experiments. The strong dark matter self interactions may ameliorate small-scale structure problems, while the strongly first-order phase transition may be confirmed at future gravitational wave observatories.
\end{abstract}

\maketitle

\section{Introduction}

Two  of  the greatest mysteries in physics are the origin of the baryon asymmetry and the nature of dark matter. The first is puzzling because, although baryon and lepton number are individually conserved at tree level in the Standard Model (SM), cosmological measurements observe a net baryon asymmetry. 

One historically popular area of study is electroweak baryogenesis \cite{Kuzmin:1985mm,Cohen:1990py,Cohen:1990it,Turok:1990in,Turok:1990zg,McLerran:1990zh,Dine:1990fj,Cohen:1991iu,Nelson:1991ab,Cohen:1992yh,Farrar:1993hn}, in which the baryon asymmetry arises from a strongly first-order electroweak phase transition. Because the minimal SM phase transition is a crossover \cite{Kajantie:1995kf,Kajantie:1996qd,Rummukainen:1998as} and CP-violation is too small \cite{Jarlskog:1985ht,Gavela:1993ts,Gavela:1994dt,Huet:1994jb}, models typically introduce additional singlet scalars \cite{Dine:1990fj,Espinosa:2011eu} or an extra Higgs doublet \cite{Turok:1990in,Turok:1990zg,McLerran:1990zh,Cohen:1991iu,Nelson:1991ab,Cohen:1992yh,Cline:1995dg, Cline:1996mga,Fromme:2006cm}. However, strong constraints on SM CP-violation have made these models increasingly in tension with experiment~\cite{Andreev:2018ayy}.

The SM must also be extended to account for dark matter, which observations~\cite{Aghanim:2018eyx} show to be roughly five times as abundant as visible matter. The similarity of dark and baryon abundances has motivated studies of asymmetric dark matter, in which the baryon and dark matter asymmetries originate from the same mechanism (see the classic reviews~\cite{Davoudiasl:2012uw,Petraki:2013wwa,Zurek:2013wia} and references therein).

In this \textit{Letter}, we introduce a minimal model in which the baryon and dark matter asymmetries originate from electroweak-like baryogenesis in a dark sector with an $\su3\times\su2\times\U1$ gauge group, two Higgs doublets, and one generation of SM-like matter content. 
A right-handed neutrino singlet and the SM electroweak sphaleron transfer the dark lepton asymmetry into an SM baryon asymmetry.
The symmetric component of the dark baryons annihilates into massive dark photons, which decay to SM states via a testable kinetic mixing, leaving the asymmetric component as $\GeV$-scale hadronic dark matter.

We consider two dark matter possibilities in detail. In the first, the dark antineutron is the lightest baryon and comprises all of dark matter. In the second, the dark antiproton is the lightest baryon and acts as dark matter together with dark pions. We find both of these scenarios are testable at current and future dark photon and direct detection experiments. Because the dark matter consists of $\GeV$-scale dark hadrons, they may also have velocity-dependent self-interactions at the correct scale to address small-scale structure issues~\cite{Chu:2018fzy}.

This paper builds upon recent work \cite{Hall:2019ank} on electroweak-like baryogenesis in a dark sector with an $\su2$ gauge group and two ``lepton" doublets. 
There is an extensive history of dark sectors with an $\su3\times\su2\times\U1$ gauge group, particularly in the context of mirror world models (see~\cite{Berezhiani:2003xm,Foot:2004pa} for a review and~\cite{Ibe:2018juk,Ibe:2018tex,Ibe:2019ena} for recent interesting examples).
The idea that the SM baryon asymmetry is the result of a dark phase transition (``darkogenesis'') was originally proposed in Ref.~\cite{Shelton:2010ta} and developed in {\it e.g.}\/, \cite{Dutta:2010va,Petraki:2011mv,Walker:2012ka,Servant:2013uwa}.
However, whereas darkogenesis models typically rely on higher-dimensional operators or a messenger sector in order to transfer the baryon asymmetry to the SM, we use a neutrino portal in a minimal, renormalizable model. Ref~\cite{Gu:2017rzz} considered a model with an $\su3 \times \su2_R \times \U1$ dark gauge sector and mirrored particle content. However, whereas they relied on a singlet scalar and CKM-like baryogenesis, we use the two Higgs doublet mechanism.

\section{Dark-Sector Baryogenesis}

The dark sector contains an $\su3'\times\su2'\times\U1'$ gauge group, one SM-like matter generation including a right-handed singlet neutrino $\bar{n}'_\RH$, and two Higgs doublets $\Phi_{\{1,2 \}}$,
\begin{equation}
    Q',u'_\RH,d'_\RH,L',e'_\RH,\bar{n}'_\RH, \, \Phi_1, \Phi_2.
\end{equation}
Throughout this \textit{Letter}, superscripts $'$ on SM particles refer to their dark-sector counterparts.
Unlike the SM mass hierarchy, we assume that dark leptons are heavy while quarks are light.

The dark gauge sector is directly analogous to the SM with the exception that the the dark $\U1'_{\rm EM}$ photon is massive and dark hypercharge kinetically mixes with SM hypercharge. After electroweak symmetry breaking in both the dark and SM sectors, these features may be parameterized in terms of the dark photon as
\begin{equation}
	\label{eq:kinmix}
	\mathcal{L} \supset \frac{\epsilon}{2} F_{\mu\nu} F'^{\mu\nu} + \frac{1}{2} m_\gamp^2 A'_\mu A'^\mu.
\end{equation}
The right-handed neutrino singlet is coupled to both dark sector Higgses as well as the SM Higgs
\begin{equation}
    \mathcal{L} \supset Y_n^a \bar{L}' \tilde{\Phi}_a \bar{n}'_\RH + y_N \bar{L} \tilde{H} \bar{n}'_\RH+c.c.,
\end{equation}
for $a \in \{1,2\}$ where $\tilde{H} = i \sigma_{2} H^{*}$ and similarly for $\tilde{\Phi}_a$. Each particle may possess distinct Yukawa couplings $Y^a$ to the two Higgs doublets. 
It is also possible that the dark neutrino has a Majorana mass term. This term would violate dark lepton number and could suppress the final asymmetries; so we conservatively assume the Majorana mass is less than a keV.

Dark-sector baryogenesis proceeds through the two-Higgs doublet mechanism at the dark electroweak phase transition, which we assume occurs before the SM electroweak crossover.
This results in a $B'+L'$ asymmetry that is primarily driven by the dark electron, which we take to have an $\order1$ Yukawa coupling.
The precise parameter space over which the two-Higgs doublet mechanism may generate a sufficient baryon asymmetry has been the subject of extensive study; see {\it e.g.} \cite{Turok:1990in,Turok:1990zg,McLerran:1990zh,Cohen:1991iu,Nelson:1991ab,Cohen:1992yh,Cline:1995dg, Cline:1996mga,Fromme:2006cm}. Baryogenesis with two Higgs doublets favors light Higgs masses and large quartic couplings, and in the context of extensions of the SM Higgs sector, this can cause issues such as Landau poles; together with recent electric dipole measurements \cite{Andreev:2018ayy}, this leads to significant constraints on the parameter space over which baryogenesis may occur. However, in the present setup, electroweak baryogenesis is easier to realize. Among other things, leptons diffuse further into the symmetric phase and do not suffer from suppression by the strong sphalerons~\cite{deVries:2018tgs}. Besides, EDM constraints do not apply and the parameter space in the dark scalar sector is almost entirely unconstrained. Hence, we expect that it should not be difficult to achieve the required baryon asymmetry and will not perform an in-depth analysis in this paper.

The aftermath of the dark first-order phase transition will produce a gravitational wave signal that could fall in the detection range of future gravitational wave observatories such as LISA, BBO, and DECIGO. Because the spectrum is generic and our model space is so unconstrained, we will not perform an in-depth investigation of gravitational wave signals here; see \cite{Hall:2019ank} for the expected spectrum.

Following dark-sector baryogenesis \cite{Hall:2019ank}, the dark leptons are in equilibrium with the SM through the neutrino portal via the process $\Phi' \nu' \leftrightarrow H \nu$ and $\bar{n}'_\RH \leftrightarrow H \nu$. Together with the SM sphaleron, this will transfer some of the initial dark lepton asymmetry into an SM baryon and lepton asymmetry. At some temperature, the remaining leptons will decay to the SM through the processes $e' \rightarrow \nu' \bar{u}' d '$ and $\nu' \rightarrow \nu H$, leaving only quarks and photons in the dark sector. Following hadronization, the symmetric component of the dark baryons will annihilate into dark photons (through {\it e.g.}\/, $\pi'^+ \pi'^- \rightarrow \gamma' \gamma'$ and $\pi'^+ \pi'^- \rightarrow \pi'^0 \pi'^0$, $\pi'^0 \rightarrow \gamma' \gamma'$) which in turn decay into the SM, leaving only the asymmetric baryon number in the dark sector	.

If the dark neutrino decays after the SM sphaleron has decoupled, we find SM baryon and lepton asymmetries \cite{Hall:2019ank}
\al{
	B = - \frac{36}{133} B' \, , \quad 
	L = \frac{97}{133} B' \, .
	\label{eq:light_crossover}
}
If the dark neutrino is heavy and decays before the SM sphaleron has decoupled, we find
\begin{equation}
	B = -\frac{12}{37} B' \, , \quad 
	L = \frac{25}{37} B'
	\label{eq:heavy_crossover}
\end{equation}
The asymmetries will differ if the SM electroweak phase transition is strongly first-order instead of a crossover and may be found in Ref.~\cite{Hall:2019ank}.

This model generalizes to one with a fully mirrored dark sector. The three families of dark quarks could motivate GUT-scale equivalence of the $\su3$ and $\su3'$ gauge couplings which could follow similar RG flows to the IR and explain the coincidence of the dark and SM matter densities.

\section{Dark Matter and Experimental Signatures}

Following dark-sector baryogenesis, the remaining asymmetric hadronic content will be dark matter and is overall neutral due to individual conservation of both dark and SM $\U1_{\rm EM}$ charges. Below the confinement scale of $\su3'$, $\Lambda_{\su3'}$, the only remaining dark sector particles are hadrons and photons. Since the dark quark masses do not affect the baryogenesis mechanism (as long as their Yukawas are sufficiently smaller than the much heavier dark leptons), there are different viable dark matter scenarios.

Since the dark leptons are heavy, much of the dark hadronic symmetric entropy density is transferred into $\pi'^0$. To ensure the dark sector does not overclose the universe, we  require $\pi'^0$ to decay. This is easily achieved through the dominant decay mode to two dark photons as long as $m_\gamp \le m_{\pi'_0}/2$. In order for the dark photon to decay into the SM, we require $m_\gamp \ge 2m_e$. The decay rate of the dark photon to a pair of SM leptons is
\al{
\Gamma_{\gamp \to \bar{l} l} = \frac{\alpha \epsilon^2 \prn{m_\gamp^2 + 2m_l^2}}{3 m_\gamp} \sqrt{1-\frac{4 m_l^2}{m_\gamp^2}}.
}
For dark photon masses below a GeV, the decay rate into hadronic channels is non-perturbative. We infer the decay rates from the branching ratios derived from measured ratios of hadronic final-state cross sections to those of muons in $e^+ e^-$ collisions~\cite{Liu:2014cma}. We require the resulting total decay rate of the dark photon to be faster than Hubble before SM neutrinos decouple around $T \sim 3 \MeV$~\cite{Mangano:2006ar}, which is true for all dark photon masses we consider as long as $\epsilon \gtrsim \order{10^{-10}}$.

With these common considerations outlined, we discuss two distinct limits: one in which all of dark matter is the dark antineutron, $\bar{n}'$, and the other in which dark matter is comprised of equal numbers of dark antiprotons, $\bar{p}'$, and $\pi'^+$, assuming $|m_{\bar{n}'}-m_{\bar{p}'}| \gtrsim 100$~MeV.

\subsection{Dark Neutron Dark Matter}

\begin{figure*}[t!]
\includegraphics[width =0.7\columnwidth]{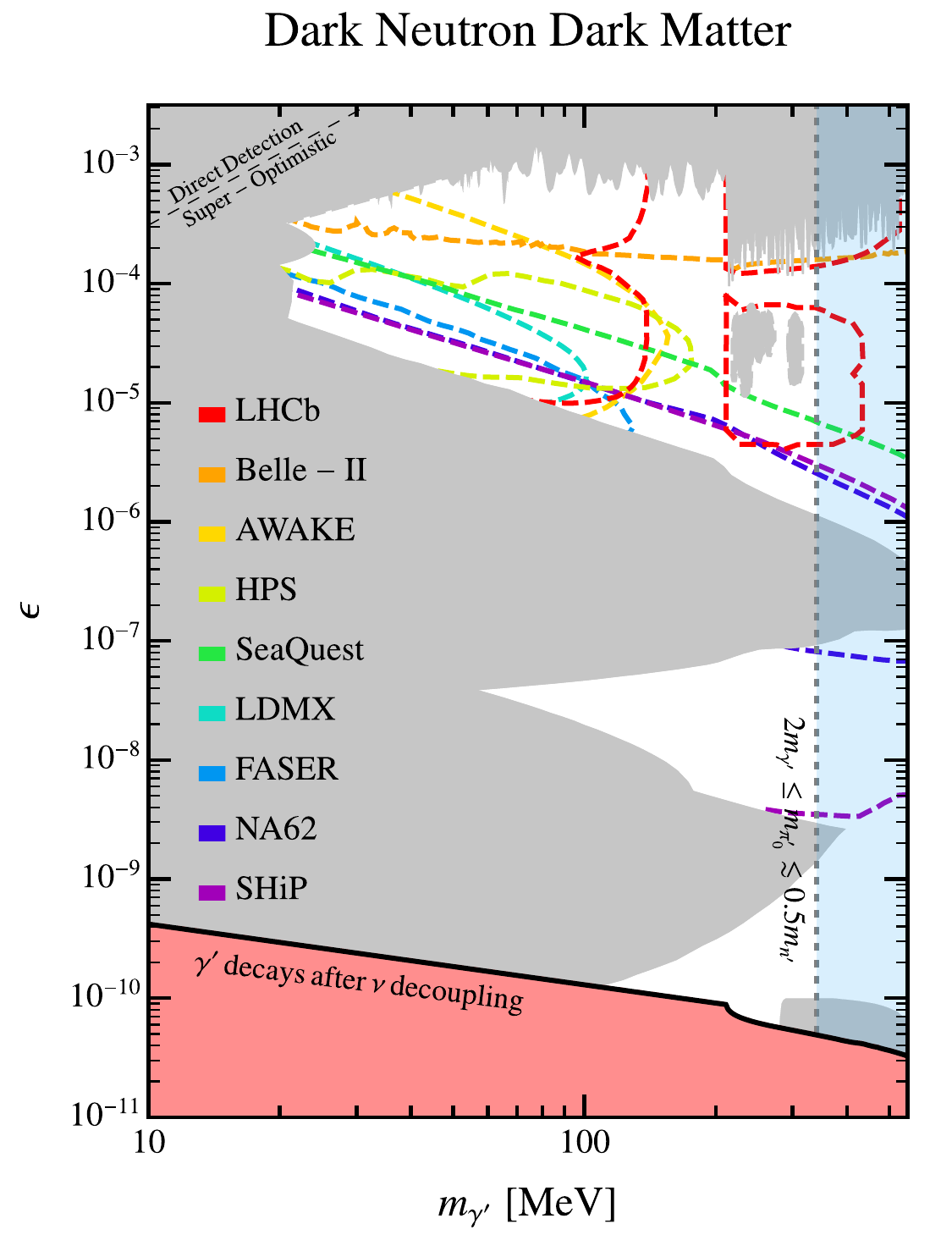} 
\hspace{0.3\columnwidth}
\includegraphics[width =0.7\columnwidth]{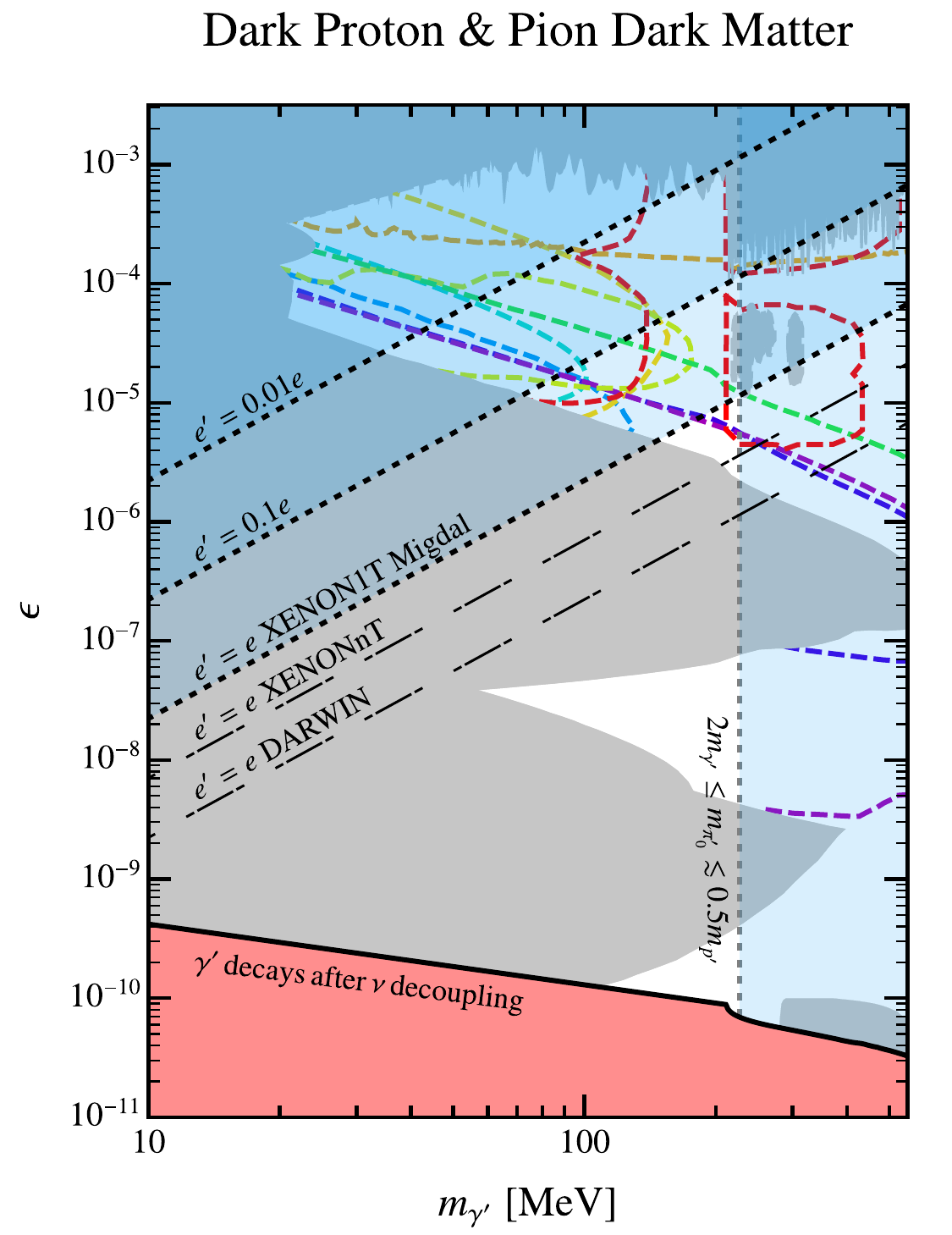}
\caption{\label{fig:darkphoton} Viable dark photon parameter space for asymmetric dark antineutron (\textbf{left}) and dark antiproton and pion (\textbf{right}) dark matter. Existing constraints on dark photons from  experiments~\cite{Alexander:2016aln,Banerjee:2018vgk,Aaij:2017rft,Aaij:2019bvg,Parker:2018vye},  supernovae~\cite{Chang:2016ntp,Hardy:2016kme}, and BBN~\cite{Fradette:2014sza} are dark gray. The constraints specific to our models from dark pion and dark photon decays are in light blue and red, respectively. Rainbow colors show projections from future experiments~\cite{Ilten:2015hya,Ilten:2016tkc,TheBelle:2015mwa,Alexander:2016aln,Caldwell:2018atq,Celentano:2014wya,Berlin:2018pwi,Berlin:2018bsc,Ariga:2018uku,NA62:2312430,Alekhin:2015byh}. Direct detection constraints from XENON1T~\cite{Aprile:2019jmx} for various $e'/e$ are shown with dashed lines, as are na\"ive projections for XENONnT~\cite{Aprile:2018dbl} and DARWIN~\cite{Aalbers:2016jon}.} 
\end{figure*}

In this scenario, the lightest dark baryon is the antineutron with $m_{\bar{p}'} - m_{\bar{n}'} \approx m_{u'} - m_{d'} \gtrsim 100$~MeV, while both quarks are light, $m_{u'}, m_{d'} < \Lambda_{\su3'}$.  After annihilations $\bar{p}' \, \pi'^+ \rightarrow \bar{n}' \,\gamp, \, \pi'^+ \, \pi'^- \rightarrow \pi'^0 \,\pi'^0$ and decays $\pi'^0 \rightarrow \gamp \gamp, \, \gamp \rightarrow \text{SM}$, the entire dark baryon asymmetry is in $\bar{n}'$ which forms all of dark matter.
The dark matter mass is precisely determined by the relative baryon and dark matter abundances \cite{Aghanim:2018eyx},
\al{
\frac{\Omega_{c}}{\Omega_b} = \frac{B'}{B} \frac{m_{\bar{n}'}}{m_p} = 5.364.
}
Given Eqs.~\ref{eq:light_crossover},~and~\ref{eq:heavy_crossover}, we predict a dark matter mass \footnote{These predictions are subject to calculable $\alpha_s/\pi$ corrections in chemical equilibrium at the percent level.}
\begin{align}
\label{eq:mnDlightcross}
    m_{\bar{n}'} &= 1.36\GeV \quad (N_R' \text{ light})
    \\
     m_{\bar{n}'} &= 1.63\GeV \quad (N_R' \text{ heavy}).
\end{align}
Although the $\bar{n}'$ is neutral, it should possess a magnetic moment similar to that of the SM neutron. This, combined with the $\gamma'$-$\gamma$ kinetic mixing, allows $\bar{n}'$ to scatter off protons in nuclei with a cross section \footnote{Both the dark antineutron charge radius and the possible Higgs portal give subdominant contributions to this scattering.}
\al{
\label{eq:sigmanDMp}
\sigma_{\bar{n}'p} \! \approx \! \epsilon^2 e^2 e'^2 F_2^{\bar{n}'2} v^4 \! \frac{m_p^4 m_{\bar{n}'}^2 \! \prn{3m_p^2+2m_p m_{\bar{n}'}+5m_{\bar{n}'}^2}}{6 \pi m_\gamp^4 \prn{m_p+m_{\bar{n}'}}^6},
}
where $v$ is the incoming dark matter velocity and $F_2^{n}\approx -1.913$ for the SM neutron. The most stringent spin-independent, per-nucleon cross section constraint on dark matter with masses $m_\chi \sim 1 \text{ GeV}$ comes from XENON1T~\cite{Aprile:2019jmx}. In particular, for $m_{\bar{n}'} = 1.36\GeV$, the bound on the dark matter-nucleon cross section is $\sigma^{\rm SI} < 7.6 \times 10^{-40} \text{ cm}^2$. This bound assumes equal couplings of the dark matter to neutrons and protons, but $\bar{n}'$ only scatters off protons, so the upper limit for $\bar{n}'p$ scattering is slightly larger:
\al{
\label{eq:sigmanDMpbound}
\sigma_{\bar{n}'p} \! < \! \prn{\frac{A}{Z}}^2 \! 7.6 \! \times \! 10^{-40} \text{ cm}^2 \approx  4.4 \! \times \! 10^{-39} \text{ cm}^2.
}
There are also limits on the self-interaction among dark matter particles from galaxy clusters $\sigma \lesssim 0.2 \, \mbox{cm}^2/{\rm g}$ \cite{Elbert:2016dbb,Bondarenko:2017rfu,Harvey:2018uwf}.
The neutrons in the SM have an astoundingly large cross section at low energies, $\sigma \approx 4.5 \times 10^{-23} \, \mbox{cm}^2$, much larger than the geometric cross section $\approx 10^{-25} \, \mbox{cm}^2$.  This is regarded as a consequence of accidental (and unnatural) cancellations in the effective field theory (see, {\it e.g.}\/, \cite{Kaplan:1998tg,Kaplan:1998we,Bedaque:2002mn}) and is not generic.  According to recent lattice QCD calculations from the HAL QCD collaboration \cite{Inoue:2011ai}, the self-interaction among $\bar{n}'$ is below the limit for rather heavy dark pions, $m_{\pi'} \gtrsim 0.4 m_{\bar{n}'}$.\footnote{This is still subject to uncertainties given disagreements with the NPLQCD collaboration \cite{Beane:2012vq,Beane:2013br} (with a possible resolution \cite{Iritani:2018vfn}), and the calculations are in the flavor $SU(3)$ limit.  It is also possible that much smaller $m_{\pi'}$ leads to small self-interaction, but it is currently beyond what can be studied on lattice.}  Therefore, this scenario prefers $m_{u'}\gtrsim 100$~MeV, which in turn allows for larger dark photon masses since $m_{\gamma'} < m_{\pi'_0}/2$. However, it is difficult to have a larger cross section at lower velocities to address the small-scale structure problems as shown with the effective range theory framework \cite{Chu:2019awd}.

The viable dark photon parameter space for the antineutron dark matter scenario is shown in Fig.~\ref{fig:darkphoton} ({\bf Left}) with current constraints from  experiments~\cite{Alexander:2016aln,Banerjee:2018vgk},  supernovae~\cite{Chang:2016ntp,Hardy:2016kme}, and BBN~\cite{Fradette:2014sza}, as well as the projected sensitivities of upcoming experiments including LHCb~\cite{Ilten:2015hya,Ilten:2016tkc}, Belle-II~\cite{TheBelle:2015mwa,Alexander:2016aln}, AWAKE~\cite{Caldwell:2018atq} ($10^{16}$ electrons of $50 \text{ GeV}$), HPS~\cite{Celentano:2014wya}, SeaQuest~\cite{Berlin:2018pwi}, LDMX~\cite{Berlin:2018bsc} (HL-LDMX with $E_{\rm beam}=16 \text{ GeV}$), FASER~\cite{Ariga:2018uku} (LHC Run 3 with $150 \text{ fb}^{-1}$), NA62~\cite{NA62:2312430}, and SHiP~\cite{Alekhin:2015byh}. Additionally, the NA64 bounds should improve soon~\cite{Banerjee:2019dyo}. Note also that even spectroscopy of resonance states is possible at $e^+ e^-$ colliders \cite{Hochberg:2015vrg,Hochberg:2017khi}.  The figure assumes the scenario in which $m_{\bar{n}'} = 1.36\GeV$ (cf.~\Eq{mnDlightcross}) and $u'$ prefers $m_{u’} \gtrsim 100$~MeV so that $m_{\pi'} \sim 0.5 m_{\bar{n}'}$. In addition to making the $\bar{n}'$ self-interactions consistent with constraints, this allows dark photons as heavy as $m_{\gamp} \sim 0.25 m_{\bar{n}'} = 0.34\GeV$.\footnote{This upper bound on the dark photon mass will relax at higher values of kinetic mixing because $\pi'^0 \rightarrow \gamp \gamp^* \rightarrow \gamp e^+ e^-$ would be possible, but we do not consider this further.} 

Interestingly, while there is currently decades of viable parameter space in which the dark photon mass and kinetic mixing can achieve the asymmetric dark antineutron dark matter, much of this will be probed by future experiments. Since the cross section in \Eq{sigmanDMp} is velocity-suppressed, current and future direct detection experiments are far from probing the viable dark photon parameter space. To illustrate this, we na\"ively assume the XENON1T bound in \Eq{sigmanDMpbound} scales linearly with exposure and project the constraint for XENON1T with 100 times its current exposure (as in DARWIN~\cite{Aalbers:2016jon}) as a thin dashed line in the upper left of Fig.~\ref{fig:darkphoton}. Additionally, we incorrectly assume that all incoming dark matter have the largest possible velocity $v_{\rm max} = v_{\text{esc}}+v_{\rm E} \sim \prn{550+240} \text{ km}/\text{s}$ (the sum of the escape and Earth velocities in the galactic frame). Clearly, such neutral dark matter seems well outside the current direct detection bounds and future dark photon searches will better probe the viable parameter space.

\subsection{Dark Antiproton \& Pion Dark Matter}

Next, we consider the case $m_{u'} < m_{d'} < \Lambda_{\su3'}$ so that the antiproton is the lightest dark baryon. Similar to the dark antineutron case, we assume $m_{d'} - m_{u'} \gtrsim 100 \text{ MeV}$ to guarantee that the $\bar{n}'$ abundance is negligible. Conservation of $\U1'_{\rm EM}$ charge implies an equal number of $\pi'^+$ comprising a subcomponent of dark matter. To reproduce the observed relic abundance, the masses satisfy
\al{
\label{eq:pandpionDMmasscond}
\frac{B'}{B} \frac{m_{\bar{p}'}+m_{\pi'^+}}{m_p} = 5.364.
}
Interestingly, $\bar{p}'$ and $\pi'^+$ may scatter resonantly in the $p$-wave through $\Delta^{\prime 0}$,
which may ameliorate small-scale structure problems if the resonant velocity is $v_R \sim 100 \, \mbox{km/s}$ with a constant $s$-wave cross section $\sigma/m \sim 0.1 \, \mbox{cm}^2/\mbox{g}$ \cite{Chu:2018fzy}. The possibility of this threshold resonance prefers $m_{\pi'^+}/m_{\bar{p}'} \sim 0.4$, which is also desirable to permit larger dark photon masses. The ``Coulomb" potential barrier may also lead to a $\bar{p}'\bar{p}'$ resonance in the $s$-wave.

For illustrative purposes, we pick the dark antiproton and pion masses to be $m_{\bar{p}'} = 2 m_{\pi'} = 0.908\GeV$ for the case that $\bar{n}'_\RH$ is light and $m_{\bar{p}'} = 2 m_{\pi'} = 1.09\GeV$ the case that $\bar{n}'_\RH$ is heavy.
While any masses satisfying \Eq{pandpionDMmasscond} with the baryon asymmetry ratio given by \Eq{light_crossover} are possible, a larger dark pion mass leads to a wider viable dark photon parameter space. 

For the available parameter space, the would-be Bohr radius $\alpha'/m_{\pi'^+}$ of the $\bar{p}'$-$\pi'^+$ ``atom'' is longer than the range of the dark-photon exchange force, $1/m_\gamp$, so we do not expect these atoms to form. Therefore, direct detection experiments can probe $\bar{p}'$-$p$ scattering with
\al{
\sigma_{\bar{p}' p} \approx \epsilon^2 e^2 e^{\prime 2} 
\frac{m_p^2 m_{\bar{p}'}^2}{\pi (m_p+m_{\bar{p}'})^2 m_{\gamma'}^4},
}
and similarly for $\pi'$-$p$ scattering. The XENON1T~\cite{Aprile:2019jmx} bound is quite weaker in this heavy-ish pion case since for $m_{\bar{p}'} = 0.908\GeV$, the bound on the dark matter-nucleon cross section is $\sigma^{\rm SI} < 2.0 \times 10^{-39} \text{ cm}^2$. Additionally, the upper limit for $\bar{p}'p$ (or $\pi'^+ p$) scattering is larger by $\prn{A/Z}^2$ due to the lack of coupling to neutrons.

The viable dark photon parameter space for the dark antiproton and pion dark matter scenario is shown in Fig.~\ref{fig:darkphoton} ({\bf right}) for $m_{\bar{p}'} = 2 m_{\pi'} = 0.908\GeV$.
The direct detection limit  weakens if $e'$ is much smaller than $e$. To demonstrate this, we show the XENON1T constraint assuming $e'/e=\{ 1,0.1,0.01 \} $ as dashed black contours. There is still a large viable parameter space and future improvements in the limits appear promising. We na\"ively assume that XENONnT~\cite{Aprile:2018dbl} with its larger exposure will increase the current bound by an order of magnitude, though the exact improvement in sensitivity from this Migdal effect analysis is not so obvious~\cite{Aprile:2019jmx}. We also show what DARWIN~\cite{Aalbers:2016jon} may probe with its possible additional order of magnitude improvement. Interestingly, it appears that future direct detection experiments may be competitive with and even exceed the sensitivity of dark photon experiments.

\subsection{Other Dark Hadron Dark Matter}

Yet another possibility is that there is only one light dark quark, let's say $u'$. Then, dark matter is partially comprised of a dark $\Delta'^{--} (\bar{u}' \bar{u}' \bar{u}')$ baryon whose abundance comes from the dark baryon asymmetry. There are also twice the number of dark ``pions'' $\pi'^+ (u' \bar{d}'),$ now heavier than $\Lambda_{\su3'}$. To produce the observed dark matter relic abundance, we require
\al{
\frac{B'}{B} \frac{m_{\Delta'^{--}} + 2 m_{\pi'^+}}{m_p} = 5.364,
}
where $m_{\Delta'^{--}} \approx \Lambda_{\su3'}$.  Besides the possible difference in masses, the direct detection would be similar to the $\bar{p}'$ and $\pi'^+$ dark matter case above. The symmetric component annihilates into $\eta(\bar{u}'u') \rightarrow \gamma'\gamma' \rightarrow 2(e^+ e^-)$.  We do not discuss this and other variants further.

\section{Conclusion}
We have introduced a minimal renormalizable model of asymmetric matter\emph{s} from a dark first-order phase transition which leads to a detectable gravitational wave signature. Electroweak-like baryogenesis in the dark sector generates a dark-sector asymmetry which is then ferried to the SM via a neutrino portal. Kinetic mixing between the dark and SM photons allows for the symmetric dark-sector entropy to safely transfer to the SM, while also providing a means for the remaining asymmetric dark matter to scatter in direct detection experiments. In the case of dark antineutron dark matter, we find decades of viable dark photon parameter space which will be explored in the near future by many upcoming experiments. If instead the asymmetric dark matter is comprised of dark antiprotons and pions, the parameter space is currently being tested by direct detection experiments. In the latter case, self interactions among the dark matter may also ameliorate small-scale problems such as the diversity problem.  If the energy scale of the dark first-order phase transition temperature is below 1000 TeV, we expect a gravitational wave signal at future observatories.

\begin{acknowledgments}
We thank G\'eraldine Servant for discussions at the early stages of this work.  We also thank Tetsuo Hatsuda, Takashi Inoue, Andr\'e Walker-Loud, and Emanuele Mereghetti for useful information about nucleon nucleon effective range parameters. We thank Michael Williams for alerting us to the latest LHCb dark photon constraints. The work of HM was supported by the NSF grant PHY-1915314, by the U.S. DOE Contract DE-AC02-05CH11231, by the JSPS Grant-in-Aid for Scientific Research JP17K05409, MEXT Grant-in-Aid for Scientific Research on Innovative Areas JP15H05887, JP15K21733, by WPI, MEXT, Japan, Hamamatsu Photonics, and by the Deutsche Forschungsgemeinschaft under Germany’s Excellence Strategy - EXC 2121 “Quantum Universe” - 390833306. The work of EH was supported by the NSF GRFP.
\end{acknowledgments}

\end{document}